\documentclass{article}

\usepackage{PRIMEarxiv}

\usepackage[utf8]{inputenc} 
\usepackage[T1]{fontenc}    
\usepackage{tikz}
\usepackage{amsmath,amsfonts}
\usepackage{algpseudocode}
\usepackage{algorithm}
\usepackage{array}
\usepackage[caption=false,font=normalsize,labelfont=sf,textfont=sf]{subfig}
\usepackage{textcomp}
\usepackage{stfloats}
\usepackage{url}
\usepackage{verbatim}
\usepackage{graphicx}
\usepackage{cite}
\usepackage{multirow}
\usepackage{makecell}
\usepackage{booktabs}       
\graphicspath{{Figures/}}     

\definecolor{mydarkblue}{HTML}{073b4c}

\usepackage{pgfplots}

\title{A Robust Defense against Adversarial Attacks on Deep Learning-based Malware Detectors via (De)Randomized Smoothing
}

\author{
  Daniel Gibert \\
  CeADAR, University College Dublin \\
  Dublin, Ireland \\
   \And
  Giulio Zizzo \\
  IBM Research Europe \\
  Dublin, Ireland \\
   \And
  Quan Le \\
  CeADAR, University College Dublin \\
  Dublin, Ireland \\
   \And
  Jordi Planes \\
  University of Lleida \\
  Lleida, Spain \\
}

\begin{document}
\maketitle

\begin{abstract}
Deep learning-based malware detectors have been shown to be susceptible to adversarial malware examples, i.e. malware examples that have been deliberately manipulated in order to avoid detection.
In light of the vulnerability of deep learning detectors to subtle input file modifications, we propose a practical defense against adversarial malware examples inspired by (de)randomized smoothing. 
In this work, we reduce the chances of sampling adversarial content injected by malware authors by selecting correlated subsets of bytes, rather than using Gaussian noise to randomize inputs like in the Computer Vision (CV) domain. 
During training, our ablation-based smoothing scheme trains a base classifier to make classifications on a subset of contiguous bytes or chunk of bytes. At test time, a large number of chunks are then classified by a base classifier and the consensus among these classifications is then reported as the final prediction. We propose two strategies to determine the location of the chunks used for classification: (1) randomly selecting the locations of the chunks and (2) selecting contiguous adjacent chunks.
To showcase the effectiveness of our approach, we have trained two classifiers with our chunk-based ablation schemes on the BODMAS dataset. Our findings reveal that the chunk-based smoothing classifiers exhibit greater resilience against adversarial malware examples generated with state-of-the-are evasion attacks, outperforming a non-smoothed classifier and a randomized smoothing-based classifier by a great margin.
\end{abstract}


\section{Introduction}
Nowadays, anti-malware engines frequently utilize Machine Learning (ML) techniques as part of a multilayered detection system, to defend against malware and to complement traditional signature-based and heuristic-based detection methods. In this work, we focus on static ML-based malware detectors, which refer to a ML-based malware detector trained using information about computer files obtained through static analysis, i.e. the analysis of computer programs without executing them. 
Broadly speaking, static ML-based malware detectors can be grouped in two main categories: (1) Feature-based detectors and (2) end-to-end detectors. Feature-based detectors~\cite{2018arXiv180404637A} rely heavily on domain knowledge to extract a set of features used to characterize the executables, which is time consuming and requires deep knowledge of the executable's file format and assembly code. Feature engineering is a continuous process. Malware authors continually adapt and modify their malicious code to evade being detected. Thus, new features might be required in the future and old ones might become obsolete or weaponized to evade detection~\cite{DBLP:conf/dmbd/Hu022,query_free_gibert}. As a result, recent research has been directed to build models that are able to perform their own feature extraction, i.e. deep learning-based or end-to-end detectors~\cite{DBLP:conf/aaai/RaffBSBCN18,GIBERT2022117957,DBLP:conf/iclr/KrcalSBJ18}. For instance, Raff et al.~\cite{DBLP:conf/aaai/RaffBSBCN18} introduced MalConv, a shallow CNN architecture that automatically learns features directly from raw byte inputs by performing convolutions.

End-to-end detectors, including MalConv, are trained with both benign and malicious code, learning to identify common byte patterns within these files. Malware authors are aware of this circumstance, and to avoid detection they try to disguise their malicious executables to resemble benign code, causing the detection system to misclassify the modified malicious executables as benign. These executables that have been specifically modified to evade detection by machine learning-based detectors are known as adversarial malware examples. A simple. yet effective method to evade end-to-end detectors is to inject content extracted from benign examples within the malicious executables~\cite{DBLP:conf/sp/SuciuCJ19,demetrio2021functionality}. As a result,
the "benign" byte patterns found in the adversarial malware examples might end up flipping the classification output of end-to-end detectors from malicious to benign. Furthermore, researchers have developed sophisticated attacks that inject and optimize small adversarial payloads within malicious executables in a way that the resulting adversarial malware examples are minimally modified but still evade detection. Depending on the access to the models, these attacks can be divided into white-box~\cite{DBLP:journals/corr/abs-1802-04528,demetrio2021adversarial} and black-box attacks~\cite{demetrio2021functionality,YUSTE2022102643}. 
For instance, Demetrio et al.~\cite{demetrio2021adversarial} proposed an attack that is able to evade end-to-end detectors by injecting and optimizing a small sized payload of 1024 bytes between the headers and the sections of Portable Executable files.

Given the vulnerabilities of deep learning detectors against the aforementioned functionality-preserving content manipulation attacks, we propose a robust defense mechanism against adversarial malware examples inspired by (de)randomized smoothing~\cite{DBLP:conf/nips/0001F20a}, a class of certifiably robust image classifiers which have been proposed against patch attacks. Our approach works as follows: 
(1) During training, a base classifier, $f(x)$, is trained to make classifications based on an ablated version of a given input file. This ablated version is generated by only selecting a small chunk of bytes from a given input file; the rest of the file is not used for classification. This frees our detection system of being constrained by the size of the executables without any downgrade in performance.
(2) At test time, the final classification $g(x)$ is taken as the class most commonly predicted by $f$ on a set of ablated versions of the file $x$. By doing so, our (de)randomized smoothing-based classifier will be more resilient to byte manipulation attacks as the injected adversarial payloads will only affect a small portion of the executable files, and thus, won't be able to flip its classification output.

The main contributions of this work are the following:
\begin{itemize}
    \item We propose a robust model agnostic defense against functionality-preserving content manipulation attacks.
    \item We introduce two chunk-based ablation schemes specifically designed for the task of malware detection.
    \item We present an empirical evaluation of state-of-the-art deep learning malware detection models state-of-the-art evasion attacks on the BODMAS dataset, showing that the proposed smoothing scheme is more robust to such attacks compared to a baseline classifier.
\end{itemize}

The rest of the paper is organized as follows. Section~\ref{sec:related_work} provides an overview of the functionality-preserving attacks specifically designed against deep learning-based malware detectors and the defenses developed so far. Section~\ref{sec:derandomized_smoothing} presents our (de)randomized smoothing approach to defend against adversarial malware examples and introduces two chunk-based schemes specifically designed for the task of malware detection. Section~\ref{sec:evaluation} evaluates the proposed defense mechanism against various state-of-the-art evasion attacks. Finally, Section~\ref{sec:conclusions} summarizes our concluding remarks and presents some future lines of research.

\section{Related Work}
\label{sec:related_work}
In this section, we provide background on the task of malware detection and introduce the state-of-the-art evasion attacks specifically designed to evade end-to-end malware detectors.

\subsection{The Task of Malware Detection}
Malware detection refers to the task of determining if a particular piece of software is benign or malicious. Lately, deep learning-based malware detectors~\cite{DBLP:conf/aaai/RaffBSBCN18,GIBERT2022117957,DBLP:conf/iclr/KrcalSBJ18} have been proposed to detect malware based on the raw byte sequences of Portable Executable files. This approaches, instead of relying in manual feature engineering, they learn patterns directly from the raw bytes through one or more convolutional layers. As shown in the literature~\cite{DBLP:journals/corr/SzegedyZSBEGF13,DBLP:conf/pkdd/BiggioCMNSLGR13}, however, machine learning-based models are vulnerable to adversarial examples, i.e. input examples that have been deliberately manipulated to evade detection, and deep learning-based models for malware detection are not an exception~\cite{demetrio2021adversarial,demetrio2021functionality,DBLP:conf/sp/SuciuCJ19,YUSTE2022102643}.

In the context of malware detection, given a target malware detector $f$, the goal of an adversarial attack on a malicious example $x$ is to generate an adversarial example $x_{adv}$ by manipulating $x$ in such a way that $x_{adv}$ has the same functionality as $x$, but it is misclassified as benign by the target malware detector $f$. Differently from adversarial attacks on image classifiers, the adversarial malware examples do not necessarily have to look imperceptibly similar to the original malware example, but only require to keep the functionality of the executable intact. Nevertheless, the existing literature~\cite{DBLP:conf/sp/SuciuCJ19,DBLP:journals/corr/abs-1802-04528} has proposed to limit the number of introduced perturbations in the adversarial malware examples to minimize the amount of change made to the original code while still achieving the desired evasion result.

\subsection{Functionality-Preserving Content Manipulation Attacks}
\begin{figure}[ht]
    \includegraphics[width=0.5\columnwidth]{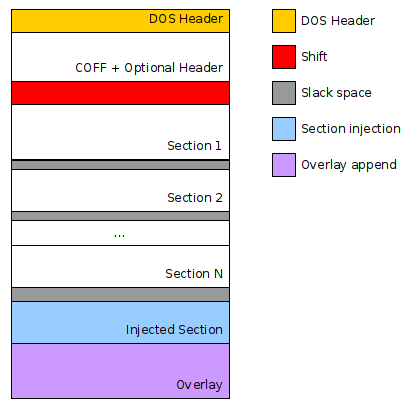}
    \centering
    \caption{A graphical depiction of the PE file format and some practical manipulations~\cite{DBLP:conf/sp/SuciuCJ19,demetrio2021adversarial,demetrio2021functionality,YUSTE2022102643}.}
    \label{fig:visual_representation_content_manipulation_attack}
\end{figure}

The attacks specifically designed in the literature to evade deep learning-based malware detectors achieve evasion by manipulating Portable Executable files by injecting new content or patching/modifying existing content within the files. This can be done by injecting the adversarial payloads into newly-created sections~\cite{demetrio2021functionality}, at the overlay or end of the file~\cite{DBLP:conf/sp/SuciuCJ19}, between the header and the sections~\cite{demetrio2021adversarial} or between sections~\cite{YUSTE2022102643}. See Figure~\ref{fig:visual_representation_content_manipulation_attack} for a visual representation of the aforementioned practical content manipulation attacks. For a more detailed description of each of the attacks we refer the readers to~\cite{DBLP:conf/sp/SuciuCJ19,demetrio2021adversarial,demetrio2021functionality,YUSTE2022102643}

A simple, yet effective technique to craft these adversarial payloads is to use benign content. This is because end-to-end malware detectors are trained with both benign and malicious executables and thus, injecting a substantial amount of benign content can potentially lead to the confusion of the classifier. In addition to the simple benign content injection attack, more elaborated evasion attacks have been recently proposed in the literature to evade end-to-end malware detectors~\cite{DBLP:conf/sp/SuciuCJ19,demetrio2021adversarial,demetrio2021functionality,YUSTE2022102643}. 

These evasion attacks can be broadly divided into white-box attacks~\cite{DBLP:journals/corr/abs-1802-04528,DBLP:conf/sp/SuciuCJ19,demetrio2021adversarial} and black-box attacks~\cite{demetrio2021functionality,YUSTE2022102643}, depending on the malware authors' access and knowledge of the target malware detector. On the one hand, white-box attacks rely on complete knowledge of the target malware detector, i.e. training algorithm, input and output, and access to the model parameters. For example, Freuk et al.~\cite{DBLP:journals/corr/abs-1802-04528} and Suciu et al.~\cite{DBLP:conf/sp/SuciuCJ19} adapted the Fast Gradient Method (FGM) originally described in~\cite{DBLP:journals/corr/GoodfellowSS14} to generate a small-sized adversarial payload that caused the malicious executables to be misclassified as benign.
Conversely, in black-box attacks, the attackers have limited information about the ML-based system that they want to attack. In a black-box setting, the attackers might have access to the inputs and outputs of the model but they have no access to information about its architecture, inner workings and parameters. Black-box attacks only require the score (score-based attacks) or the label (label-based attacks) predicted by the malware detector. 
For instance, Demetrio et al.~\cite{demetrio2021adversarial,demetrio2021functionality} proposed RAMEN and GAMMA, respectively. RAMEN~\cite{demetrio2021adversarial} is a general framework for developing adversarial attacks on ML-based malware detection systems that rely on static code analysis. 
This framework includes three novel attacks, Full DOS, Extend, and Shift, which manipulate the DOS header, extend it, and shift the content of the first section, respectively. These attacks are shown to outperform existing ones in both white-box and black-box scenarios, achieving a better trade-off in terms of evasion rate and size of the injected payload. 
Conversely, GAMMA (Genetic Adversarial Machine Learning Malware Attack) generates adversarial malware examples by injecting an adversarial payload into newly-created sections and optimizing the payload with genetic algorithms. Similarly, Yuste et al.~\cite{YUSTE2022102643} dynamically extended and optimized the bytes in unused blocks, referred to as code caves, from Portable Executable (PE) files to generate the adversarial malware examples.

\subsection{Adversarial Defenses}
As far as we are aware, two defenses have been proposed so far to defend against evasion attacks. 


Lucas et al.~\cite{advtrain:sec2023} explored the application of adversarial training to improve the robustness of deep learning-based malware detectors against certain adversarial attacks. Their findings indicate that data augmentation alone does not deter state-of-the-art attacks. However, in specific cases, models can be made slightly more resilient to attacks by adversarially training them using lower-effort or less sophisticated versions of the same attacks. For instance, in their work, Lucas et al.~\cite{advtrain:sec2023} adversarially trained a deep learning-based detector with adversarial examples generated by three attacks: (1) In-Place Replacement attack (IPR)~\cite{10.1145/3433210.3453086}, (2) Displacement attack (Disp)~\cite{10.1145/3433210.3453086} and (3) Padding attack~\cite{DBLP:conf/sp/SuciuCJ19}. The model was then subsequently evaluated against an attack that combines In-Place Replacement and Displacement. In their evaluation, the adversarially trained model exhibited an accuracy of 49\% whereas the accuracy of the original model was 25\%. Thus, their work shows that although adversarial training helps building more resilient models against the attacks seen during training, it is not enough to defend against state-of-the-art attacks, leaving the detectors vulnerable to unknown threats. In addition, generating adversarial malware examples is computationally expensive and non-trivial, and as a result, training models with adversarial training requires more computational resources compared to standard training procedures.

Alternatively, Gibert et al.~\cite{gibert2023practical} proposed a randomized smoothing scheme to improve the robustness of end-to-end malware classifiers against adversarial malware examples. Randomized smoothing is a technique that increases the robustness of ML-based classifiers against adversarial attacks by adding random noise to the input data during both training and inference, with the goal of making the classifier's predictions stable and resilient to small perturbations.
In their work, a base malware classifier $f$ is trained to make classifications based on an ablated version $\tilde{x}$ of a given executable file $x$, where $\tilde{x}$ consists of a copy of the original executable $x$ with the bytes ablated given a probability $p$. Afterwards, at inference time, $L$ ablated versions of an input executable $x$ are generated and the final classification is determined as the class that the malware classifier most frequently predicts across the set of ablated versions of the executable. Unlike adversarial training, smoothing-based classifiers focus on blurring the decision boundaries within the input space, improving the classifier's resilience against a broader range of attacks~\cite{DBLP:conf/icml/CohenRK19}. 

Randomized smoothing, however, while effective against certain types of adversarial attacks, has limitations when applied to defending against adversarial malware examples. This is because the constraints imposed by executable files present unique challenges and thus, the adversarial attacks in the malware domain differ from attacks/perturbations in image or text data. 
Specifically, in the context of malware detection, attackers often inject an adversarial payload within specific parts of the executable files rather than making arbitrary modifications to bytes. Unlike image pixels or text characters, the bytes within executable files serve functional roles, and arbitrary modifications to these bytes can break the intended functionality of the executable. As a result, randomized smoothing, which inject random noise, might not be the most suitable defense against adversarial malware examples.

\section{(De)Randomized Smoothing for Malware Detection}
\label{sec:derandomized_smoothing}
To defend against adversarial malware examples, in this paper
we propose a chunk-based ablation smoothing scheme that takes into account the particularities of the adversarial attacks specifically designed against deep learning-based malware detectors and the structured nature of Portable executable files. Our approach is inspired by (de)randomized smoothing~\cite{DBLP:conf/nips/0001F20a}, a class of certifiably robust image classifiers which have been proposed against patch attacks. 

In our (de)randomized smoothing scheme, a base malware classifier, $f$, is trained to make classifications based on an ablated version $\tilde{x}$ of a given executable file $x$. This ablated version $\tilde{x}$ consists of only a subset of contiguous bytes or chunk of bytes from the original file $x$. The rest of the file is ablated. This functionality is implemented by the operation $\text{ABLATE\_train}(x,p)$, where $p$ indicates the percentage of byte values from the original file that the sampled chunk of bytes must have. The training procedure is defined in Algorithm~\ref{alg:smoothed_classifier_training}.

\begin{algorithm}
\caption{Smoothed classifier training procedure}\label{alg:smoothed_classifier_training}%
\begin{algorithmic}
\Require training dataset $D_{train}$, malware detector $f$ with parameters $\theta$, file portion $p \in \mathbb{R} \: ,\:  0<=p<=1$
\State $\theta \gets \text{Initialize parameters} $
\For{i=1, MAX\_EPOCHS}
    \For{$x, y \in D_{train}$}
        \State $\tilde{x} \gets \text{ABLATE\_train}(x, p)$
        \State $\tilde{y}   \gets f(\tilde{x} )$
        \State $\text{Loss} \gets \text{criterion}(y, \tilde{y} )$
        \State $\theta \gets \text{Update parameters}$
    \EndFor
\EndFor
\end{algorithmic}
\end{algorithm}

\begin{figure*} 
    \centering
    \includegraphics[width=\textwidth]{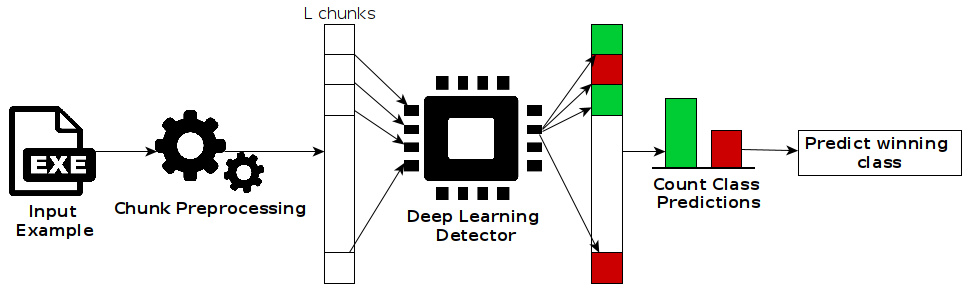}
    \caption{Illustration of the chunk-based ablation smoothing scheme. The preprocessing step extracts chunks of the input example as described in Sections~\ref{sec:randomized_ablation} and~\ref{sec:structural_ablation}}
    \label{fig:chunk_ablation_scheme_overview}
\end{figure*}

Let $x$ be an input file, $p$ be the proportion of the chunk size compared to the size of the original file, and $f$ be a base classifier. At test time, we generate $L$ chunks from $x$ using the function $\text{ABLATE\_inference}(x,p,L)$, and classify each chunk of bytes into its corresponding class using $f$. To make the final classification, we count the number of chunks of $x$ that the base classifier returns for each class and divide it by the total number of chunks $L$. An overview of the chunk-based ablation scheme is provided in Figure~\ref{fig:chunk_ablation_scheme_overview}. 

Notice that the $\text{ABLATE}$ function serves distinct purposes during training and inference phases. During training, $\text{ABLATE\_train}$ is designed to accept to two parameters: a file $x$ and a percentage value $p$. It operates by extracting a chunk from input $x$ with a size determined by the percentage $p$. 
During the test phase, the behavior of $\text{ABLATE\_inference}$ requires three input parameters: $x$, $p$, and $L$, where $L$ denotes the desired number of chunks to be extracted from $x$. At test time, the function $\text{ABLATE\_inference}$ extracts a total of $L$ chunks from the input $x$, the size of each chunk being determined by the percentage $p$.

This binary classification problem can be formally defined as follows. Let $C = \{c_1, c_2\}$ be the set of all possible classes, i.e. benign and malicious. For each class $c_i \in C$, let $N_i$ be the number of chunks of $x$ that the base classifier $f$ returns as belonging to class $c_i$. 
Then, the probability of input file $x$ belonging to class $c_i$ can be estimated as:

$$P(c_i, x) = \frac{N_i}{L}$$

where $N_i$ is the number of chunks of $x$ that belong to class $c_i$ according to the base classifier $f$.

The final classification of $x$ can then be determined as:

$$\hat{y}=argmax_{c_{i}\in C}P(c_i|x)$$

where $\hat{y}$ is the predicted class for input file $x$.


The function $\text{ABLATE\_inference}(x,p,L)$ generates $L$ chunks from the input file $x$ according to the schemes defined in Sections~\ref{sec:randomized_ablation}, and~\ref{sec:structural_ablation}. 

\subsection{Randomized Chunk-based Ablation Scheme}
\label{sec:randomized_ablation}
This section describes the Randomized Chunk-based Ablation (RCA) scheme. The RCA scheme works by dividing the input file represented as a byte sequence into several randomly sampled chunks and classifying each chunk independently. The classification results from all the chunks are then combined by a majority vote to determine the final prediction for the input file. 
This approach aims to improve the model's resiliency to adversarial examples and to reduce the sensitivity to the ordering of the input sequence.

The RCA scheme has three key components:
\begin{enumerate}
    \item Sampling $L$ chunks randomly from the input sequence using a uniform distribution.
    \item Classifying each chunk independently.
    \item Combining the classification results by taking a majority vote.
\end{enumerate}

To generate $L$ chunks from the input example, the $\text{ABLATE\_inference}(x, p, L)$ functions is used, where the parameter $p$ determines the size of the chunks. This function, defined in Algorithm~\ref{alg:randomized_smoothing_ablate_operation}, works by randomly sampling chunks from the input sequence to generate multiple versions of it.


\begin{algorithm}
\caption{Implementation of the ABLATE\_inference function for the randomized chunk-based ablation scheme.}\label{alg:randomized_smoothing_ablate_operation}%
\begin{algorithmic}
\Function{ABLATE}{$x, p, L$}
    \State chunks $\gets$ NewList()
    \State l $\gets$ Size(list)
    \State group\_size $\gets$ Ceil(l*p)
    \For{i $\gets$ 1 to L}
        \State start\_index = GenerateRandomInteger(0, l - group\_size)
        \State end\_index = start\_index + group\_size
        \State chunk\_i = x[start\_index:end\_index]
        \State AddItem(chunks, chunk\_i) 
    \EndFor
    \State \Return chunks
\EndFunction
\end{algorithmic}
\end{algorithm}

\subsection{Structural Chunk-based Ablation Scheme}
\label{sec:structural_ablation}
This section describes a different strategy for classifying byte sequences named Structural Chunk-based Ablation (SCA) scheme. In contrast to the RCA scheme, the SCA scheme samples chunks in an orderly way from the start of the input byte sequence to its end, instead of randomly sampling chunks.

The SCA scheme has three key components:
\begin{enumerate}
    \item Sampling $L$ sequentially adjacent chunks from the input sequence.
    \item Classifying each chunk independently.
    \item Combining the classification results by taking a majority vote.
\end{enumerate}

The SCA scheme takes as input a byte sequence $x$ and samples $L$ sequentially adjacent chunks of equal size according to the file percentage $p$. Notice that depending on the number of chunks to sample $L$ and the chunk size, the chunks might overlap with one another. The SCA approach is useful for classifying byte sequences where the order of the sequence is important. By sampling chunks sequentially, the SCA approach takes into account the structural properties of the input sequence, which can improve the classification accuracy.



\begin{algorithm}
\caption{Implementation of the ABLATE\_inference function for the structural chunk-based ablation scheme.}\label{alg:structural_smoothing_ablate_operation}%
\begin{algorithmic}
\Function{ABLATE}{$x, p, L$}
    \State chunks $\gets$ NewList()
    \State $l$ $\gets$ Size(list)
    \State group\_size $\gets$ Ceil($l*p$)
    \State overlap\_size $\gets$ Integer($l / L$)
    \State $a$ $\gets$ Ceil((group\_size - overlap\_size) / $L$ )
    \State $b$ $\gets$ Ceil((group\_size - overlap\_size) * $p$ )
    \State bytes\_corrected $\gets$ $b - a$
    \For{i $\gets$ 1 to L}
        \State start\_index = i $*$ (group\_size - overlap\_size - bytes\_corrected)
        \State end\_index = start\_index + group\_size
        \State chunk\_i = x[start\_index:end\_index]
        \State AddItem(chunks, chunk\_i) 
    \EndFor
    \State \Return chunks
\EndFunction
\end{algorithmic}
\end{algorithm}

\section{Evaluation}
\label{sec:evaluation}
We now empirically evaluate the proposed randomized chunk-based ablation (RCA) and structural chunk-based ablation (SCA) schemes to determine the robustness of the smoothing approaches against adversarial malware examples generated by four state-of-the-art evasion attacks.

\subsection{Experimental Setup}
Here we describe the details of the experimental setup, including data sources, machine learning models for malware detection, and parameters for the randomized and structural chunk-based ablation models.

The experiments have been run on a machine with an Intel Core i7-11800H CPU, 1xGeforce GTX3070 GPU and 32Gb
RAM. The code has been implemented with PyTorch~\cite{NEURIPS2019_9015} and will be made publicly available in our Github repository~\footnote{\url{https://github.com/danielgibert/derandomized_smoothing_for_malware_detection}} after acceptance.
\subsubsection{BODMAS Dataset}
\label{sec:bodmas_dataset}
The dataset used in this paper to evaluate the proposed ablation-based classification schemes is the BODMAS dataset~\cite{bodmas}. This dataset consists of 57,293 malware with family information (581 families and 77,142 benign Windows PE files collected from August 2019 to September 2020. Similar to \cite{gibert2023practical}, we have only considered those executables that are equal to or smaller than 1Mb. The reduced dataset consists of 39,380 and 37,739 benign and malicious executables, respectively~\footnote{Limiting the size of the executables to 1Mb has accelerated our experiments. Moreover, since some of the evaluated evasion attacks involve adding content to the executables, larger executables would need to be trimmed before being feed into the model.}. Furthermore, we have split the dataset into training (80\%), validation (10\%) and test (10\%) sets using the timestamp of each example. As a result, the training set contains the oldest examples while the test set contains the newest examples.

\subsubsection{Malware Detectors}
\begin{figure}[ht]
    \includegraphics[width=0.7\columnwidth]{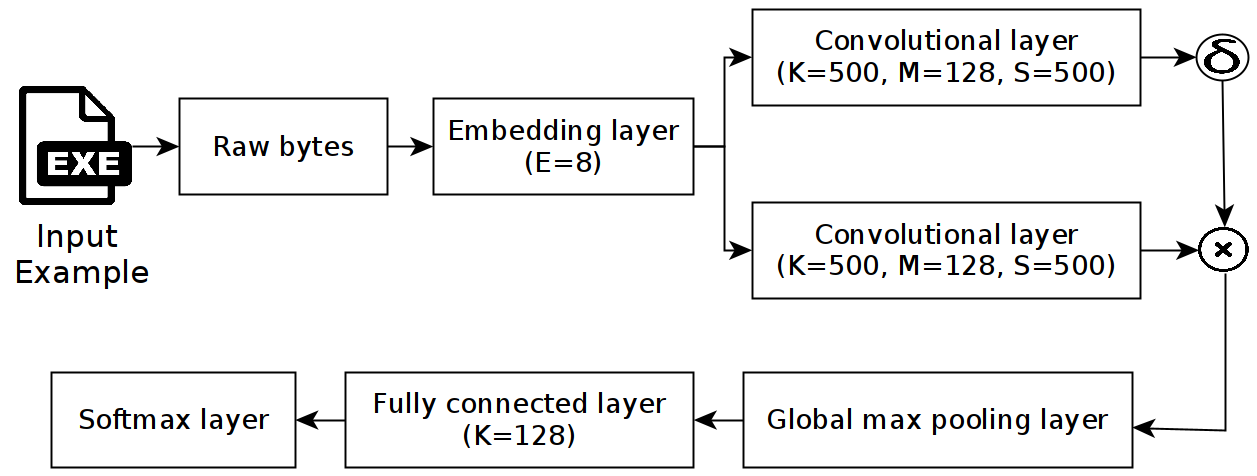}
    \centering
    \caption{Graphical depiction of the MalConv architecture.}
    \label{fig:malconv_architecture}
\end{figure}

In this work, we experiment with a deep learning-based malware detector called MalConv~\cite{DBLP:conf/aaai/RaffBSBCN18}. MalConv was one of the first end-to-end deep learning model proposed for malware detection. End-to-end models learn to classify examples directly from raw byte sequences, instead of relying on feature engineering. The network architecture of MalConv consists of an embedding layer, a gated convolutional
layer, a global-max pooling layer and a fully-connected layer. See Figure~\ref{fig:malconv_architecture}.

Using MalConv as a basis, four malware detectors have been considered:
\begin{itemize}
    \item NS-MalConv. This detector corresponds to a non-smoothed MalConv model~\cite{DBLP:conf/aaai/RaffBSBCN18}. It serves as a non-robust baseline as no specific technique has been employed to improve its robustness to evasion attacks. 
    \item RS-MalConv. This detector implements the randomized smoothing scheme proposed by Gibert et al~\cite{gibert2023practical}.
    \item RCA-MalConv. This detector implements the randomized chunk-based ablation scheme proposed in Section~\ref{sec:randomized_ablation} using MalConv as a base detector.
    \item SCA-MalConv. This detector implements the structural chunk-based ablation scheme proposed in Section~\ref{sec:structural_ablation} using MalConv as a base detector.
\end{itemize}

\subsection{Empirical Robustness Evaluation}
In this section, we empirically evaluate the robustness of the chunk-based smoothing schemes to several published evasion attacks. By doing so, we aim to provide a complete picture of the strengths and weaknesses of the proposed defenses. We provide details of the performance of the proposed defenses against non-adversarial examples in Section~\ref{sec:nonadversarial_evaluation} and we present the results of the ablation-based schemes against four state-of-the-art evasion attacks in Section~\ref{sec:sota_attacks}.

\subsubsection{Non-Adversarial Evaluation}
\label{sec:nonadversarial_evaluation}

Both randomized and structural chunk-based ablation schemes depend on two hyperparameters: (1) the proportion of bytes containing each chunk with respect to the file size, denoted by $p$, and (2) the number of chunks to use for classification, denoted by $L$. 

It is important to note that both parameters affect the behavior of both schemes differently. The greater the number of chunks used for classification in the randomized-based scheme the more stable the results produced will be. This is because the randomized-based scheme randomly samples chunks from a single example. As a result, the more chunks are sampled the less variability. On the other hand, the greater the number of chunks in the structural-based scheme the more overlapping between chunks will be. For example, if the size of the chunks is equal to 5\% the original size of the files and the number of chunks for classification is 20 then there won't be any overlapping between chunks. On the other hand, if the number of chunks for classification is 100 then there will be an 80\% of overlapping between adjacent chunks.

We would like to note that we have chosen the same hyperparameters for the MalConv model as those in the original paper~\cite{DBLP:conf/aaai/RaffBSBCN18}. The only difference between the hyperparameters of the NS-MalConv, RCA-MalConv and SCA-MalConv is the number of epochs used during training. In the case of NS-MalConv, the model is trained on the entire binary file in each step. However, for the RCA-MalConv and SCA-MalConv, only a small portion of the input file is used during each training step, and so a larger number of epochs (in this case, 50) is needed to ensure that the model is trained on enough chunks from the file. In addition, early stopping is employed, causing the training process to terminate immediately once the malware's detection validation performance drops for more than 5 epochs.

\begin{table}[ht]
\centering
\caption{Performance metrics of RCA-MalConv and SCA-MalConv models on the validation set given different chunk sizes.}
\label{tab:smoothing_performance}
\begin{tabular}{cclll}
\hline
\multirow{2}{*}{}            &                               & \multicolumn{3}{c}{\begin{tabular}[c]{@{}c@{}}Chunk size \\ (percentage of the \\original size)\end{tabular}} \\ \cline{2-5} 
                             &                               & \multicolumn{1}{c}{1\%}             & \multicolumn{1}{c}{2\%}             & \multicolumn{1}{c}{\textbf{5\%}}         \\ \hline
\multirow{2}{*}{RCA-MalConv} & \multicolumn{1}{c|}{Accuracy} & \multicolumn{1}{l|}{86.99}          & \multicolumn{1}{l|}{93.44}          & \textbf{95.66}                           \\
                             & \multicolumn{1}{c|}{F1-Score} & \multicolumn{1}{l|}{86.85}          & \multicolumn{1}{l|}{93.52}          & \textbf{95.69}                           \\ \hline
\multirow{2}{*}{SCA-MalConv} & \multicolumn{1}{c|}{Accuracy} & \multicolumn{1}{l|}{87.10}          & \multicolumn{1}{l|}{93.67}          & \textbf{95.90}                           \\
                             & \multicolumn{1}{c|}{F1-Score} & \multicolumn{1}{l|}{86.95}          & \multicolumn{1}{l|}{93.75}          & \textbf{95.93}                           \\ \hline
\end{tabular}
\end{table}

Table~\ref{tab:smoothing_performance} presents the accuracy and F1 score of the RCA-MalConv and SCA-MalConv with varying chunk sizes. It can be observed that the greatest accuracy and F1 score has been achieved by the SCA-MalConv and RCA-MalConv with a chunk size equal to 5\% of the original file size. Compared to the accuracy and F1-score of the non-smoothed classifier (Table~\ref{tab:non_adversarial_examples_performance_metrics}), the chunk-based approaches achieve similar detection accuracies compared to a non-smoothing classifier while being more robust to evasion attacks.

\begin{table}[ht]
\centering
\caption{Performance metrics of the MalConv models on the test set and the sub-test sets of 500 malware examples used for adversarial attack evaluation. With the sub-test set being composed of only malware we simply report the accuracy.}
\label{tab:non_adversarial_examples_performance_metrics}
\begin{tabular}{c c cc}
\toprule
 &  & Accuracy & F1 \\ 
\midrule
\multirow{4}{*}{Test Set} & NS-MalConv  & 98.03 & 97.98 \\
                          & RS-MalConv  & 97.61 & 97.57 \\
                          & RCA-MalConv & 97.34 & 97.35 \\ 
                          & SCA-MalConv & 97.63 & 97.64 \\ 
\midrule
 &  & \multicolumn{2}{c}{Accuracy} \\ 
\midrule
\multirowcell{4}{Test\\Sub-Set\\(500 examples)} & NS-MalConv & \multicolumn{2}{c}{97.00} \\
                          & RS-MalConv  &  \multicolumn{2}{c}{95.00} \\
                          & RCA-MalConv & \multicolumn{2}{c}{97.80} \\

                          & SCA-MalConv & \multicolumn{2}{c}{98.00}     \\  \hline

%
\bottomrule
\end{tabular}%
\end{table}


\subsubsection{Empirical Robustness Against SOTA Evasion attacks}
\label{sec:sota_attacks}
\begin{table*}[ht]
\centering
\caption{Summary of the evasion attacks used to assess the robustness of the proposed smoothing classifiers.}
\label{tab:sota_attacks}
\resizebox{\textwidth}{!}{%
\begin{tabular}{l|l|l}
\hline
Attack        & Type of Attack             & Description \\ \hline
Slack+Padding~\cite{DBLP:conf/sp/SuciuCJ19}  & White-box                &   Padds bytes at the end of the executables and manipulates the bytes from the slack space\\ \hline
Shift~\cite{demetrio2021adversarial}         & White-box     & Creates space between the headers and the sections of the executables        \\ \hline
GAMMA~\cite{demetrio2021functionality}         & Black-box       & Creates new sections with benign content          \\ \hline
Code caves~\cite{YUSTE2022102643}      & Black-box    &  Dynamically extends and optimize the size of the code caves with a genetic algorithm   \\ \hline
\end{tabular}%
}
\end{table*}

We consider four recently published attacks designed to bypass static PE malware detectors as summarized in Table~\ref{tab:sota_attacks}. To perform the attacks, we have adapted the open source implementation of the attacks from secml-malware library~\cite{demetrio2021secmlmalware}. These attacks range from manipulating very few bytes from the PE headers~\cite{demetrio2021adversarial}, to injecting hundreds to thousands of bytes~\cite{DBLP:conf/eusipco/KolosnjajiDBMGE18,DBLP:journals/corr/abs-1802-04528,DBLP:conf/sp/SuciuCJ19,demetrio2021functionality,YUSTE2022102643}. Since some attacks might take hours to run per file, we have used one smaller-sized test set containing 500 malware examples randomly subsampled from the test set.
By employing a smaller subset, we aim to reduce the computational overhead of using genetic algorithms to optimize the adversarial payloads. We would like to denote that our evaluation set is comparable in size to prior work~\cite{DBLP:conf/eusipco/KolosnjajiDBMGE18,DBLP:journals/corr/abs-1802-04528,DBLP:conf/sp/SuciuCJ19,demetrio2021adversarial}. If supported by the attack, i.e. Shift, Slack and GAMMA, early stopping is employed, causing the attack to terminate immediately once the malware detector's prediction changes from malicious to benign. 

We would like to point out that the white-box attacks cannot be directly applied against the randomized smoothing (RS-MalConv) and the (de)randomized smoothing classifiers (RCA-MalConv, SCA-MalConv) as they require computing the gradients, and the majority vote employed by the smoothing-based classifiers is not differentiable. To overcome this situation, the adversarial payloads have been optimized using genetic algorithms (GAs) as described in Gibert et al.~\cite{gibert2023practical}.

Next, the impact of the evasion attacks on the malware detectors is measured.

\subsubsection*{Slack+Padding Attack}

\begin{figure}[ht]
    \centering
    \includegraphics[width=0.6\columnwidth]{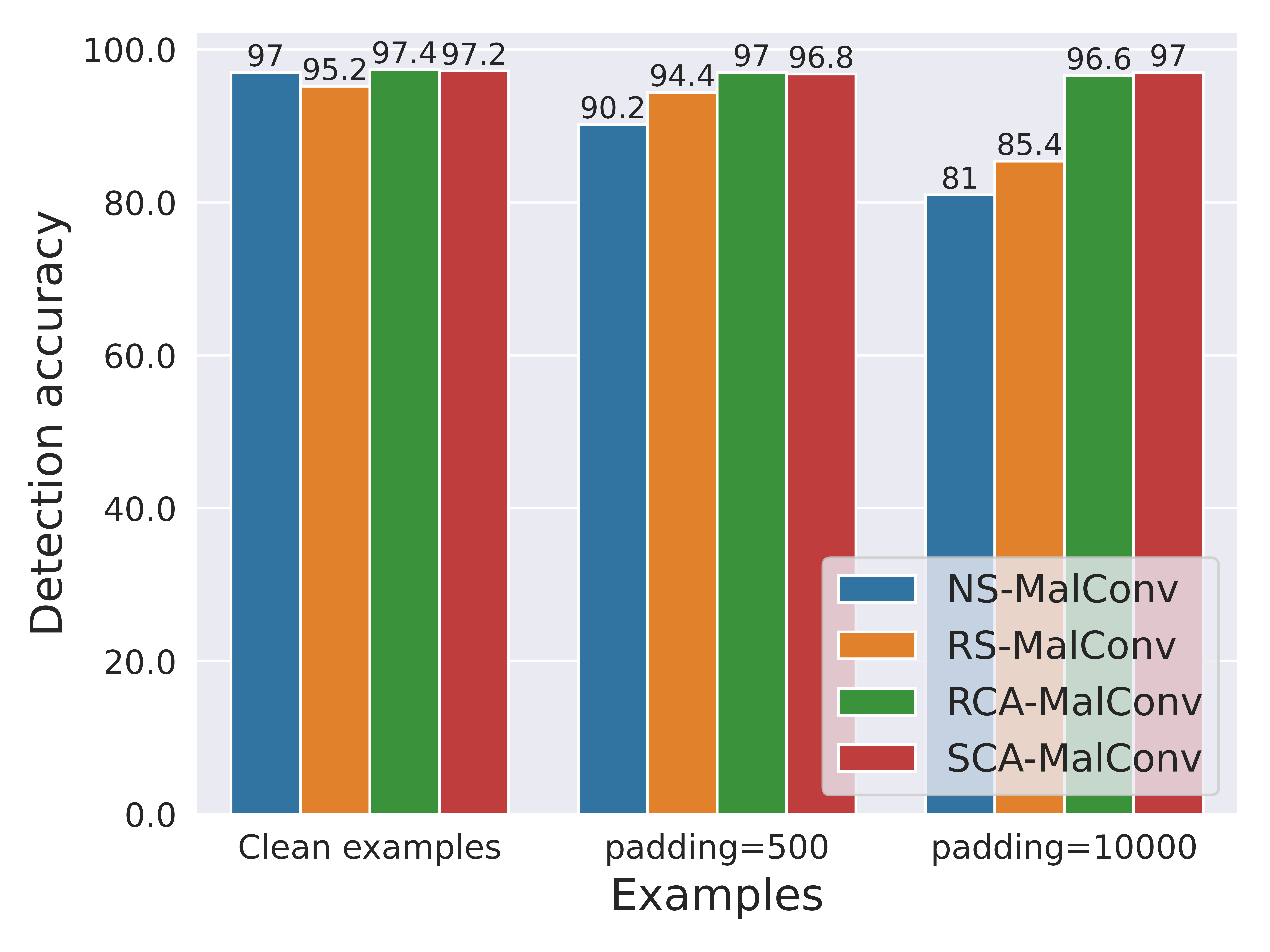}
    \caption{Detection accuracy of the malware detectors on the adversarial examples generated by Suciu et al~\cite{DBLP:conf/sp/SuciuCJ19}.}
    \label{fig:suciu_attack}
\end{figure}

Suciu et al.~\cite{DBLP:conf/sp/SuciuCJ19} proposed an evasion attack against malware detectors that modifies the bytes in the slack space and some bytes appended at the end of the file using the fast gradient sign method~\cite{DBLP:journals/corr/GoodfellowSS14} (FGSM). The FGSM method works by perturbing the input data by a small amount in the direction of the gradient of the loss function with respect to the input. The FGSM method is a fast and effective way to generate adversarial examples, as it only requires one forward and backward pass through the network to compute the gradient. Unfortunately, GAs have been used to optimize the appended payload as the FGSM method can't be applied on the smoothing-based classifiers. In our experiments, we allowed the attack to inject 500 and 10000 bytes at the end of executables. Results are shown in Figure~\ref{fig:suciu_attack}. It can be observed that the (de)randomized smoothed-based models are robust to the attack while the non-smoothed model is vulnerable to it. Notice that the SCA-MalConv achieves almost the same accuracy on both the clean and adversarial examples.

\subsubsection*{Shift Attack}
\begin{figure}[ht]
    \centering
    \includegraphics[width=0.6\columnwidth]{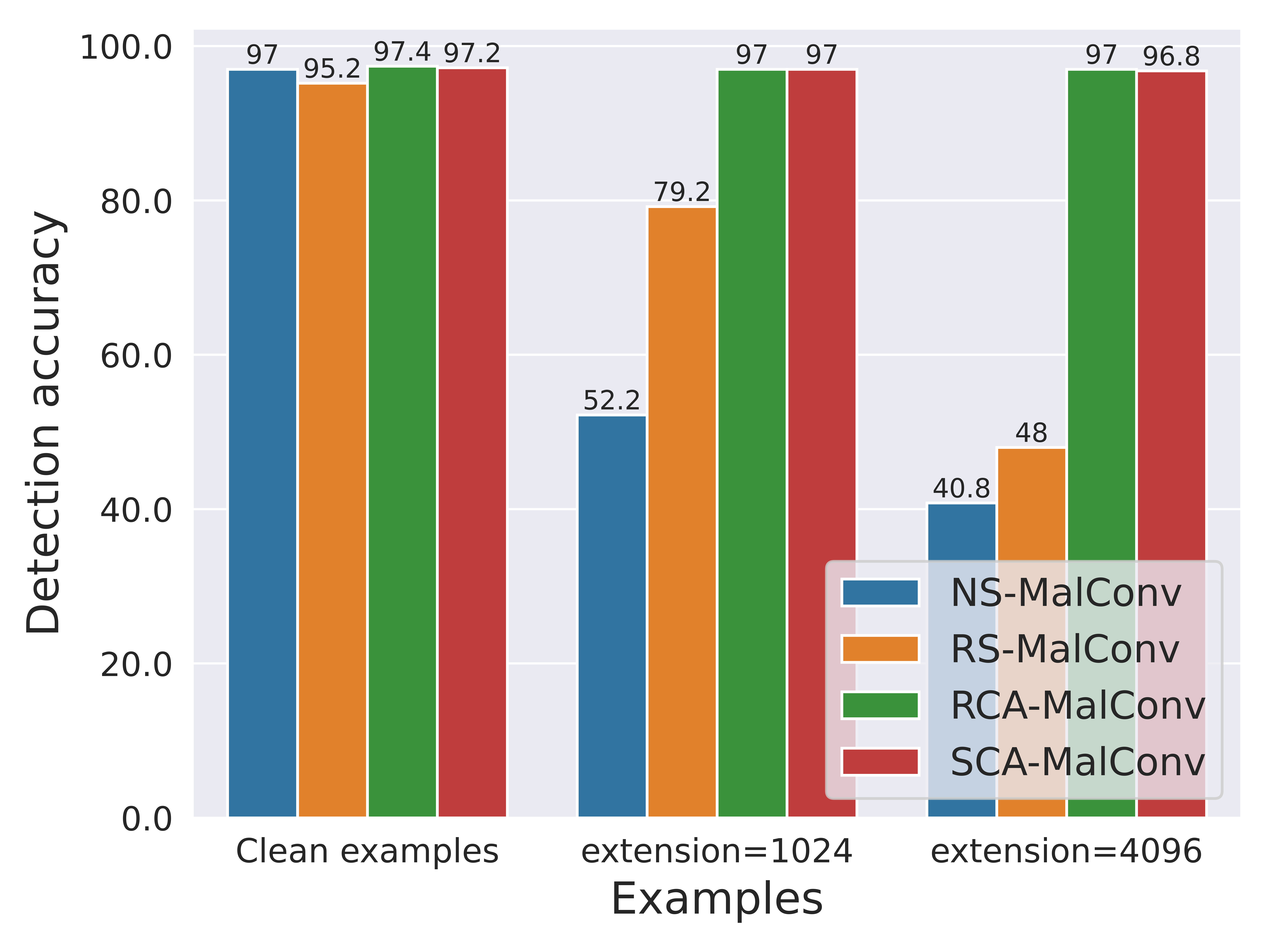}
    \caption{Detection accuracy of the malware detectors on the adversarial examples generated by the Shift attack~\cite{demetrio2021adversarial}.}
    \label{fig:shift_attack}
\end{figure}

Demetrio et al.~\cite{demetrio2021adversarial} proposed an evasion attack that generates adversarial examples by creating new space inside the executables by shifting the content of the first section and injecting there the adversarial payload. Figure~\ref{fig:shift_attack} presents the detection accuracy of the ML-based models against the shift attack for different extension amounts. Notice that when adding an adversarial payload of 4096 bytes the NS-MalConv is only able to detect 40.80\% of the adversarial examples while the RCA-MalCOnv and SCA-MalConv have 97.00\% and 96.80\% of accuracy, respectively.
\subsubsection*{GAMMA Attack}
\begin{figure}[ht]
    \centering
    \includegraphics[width=0.6\columnwidth]{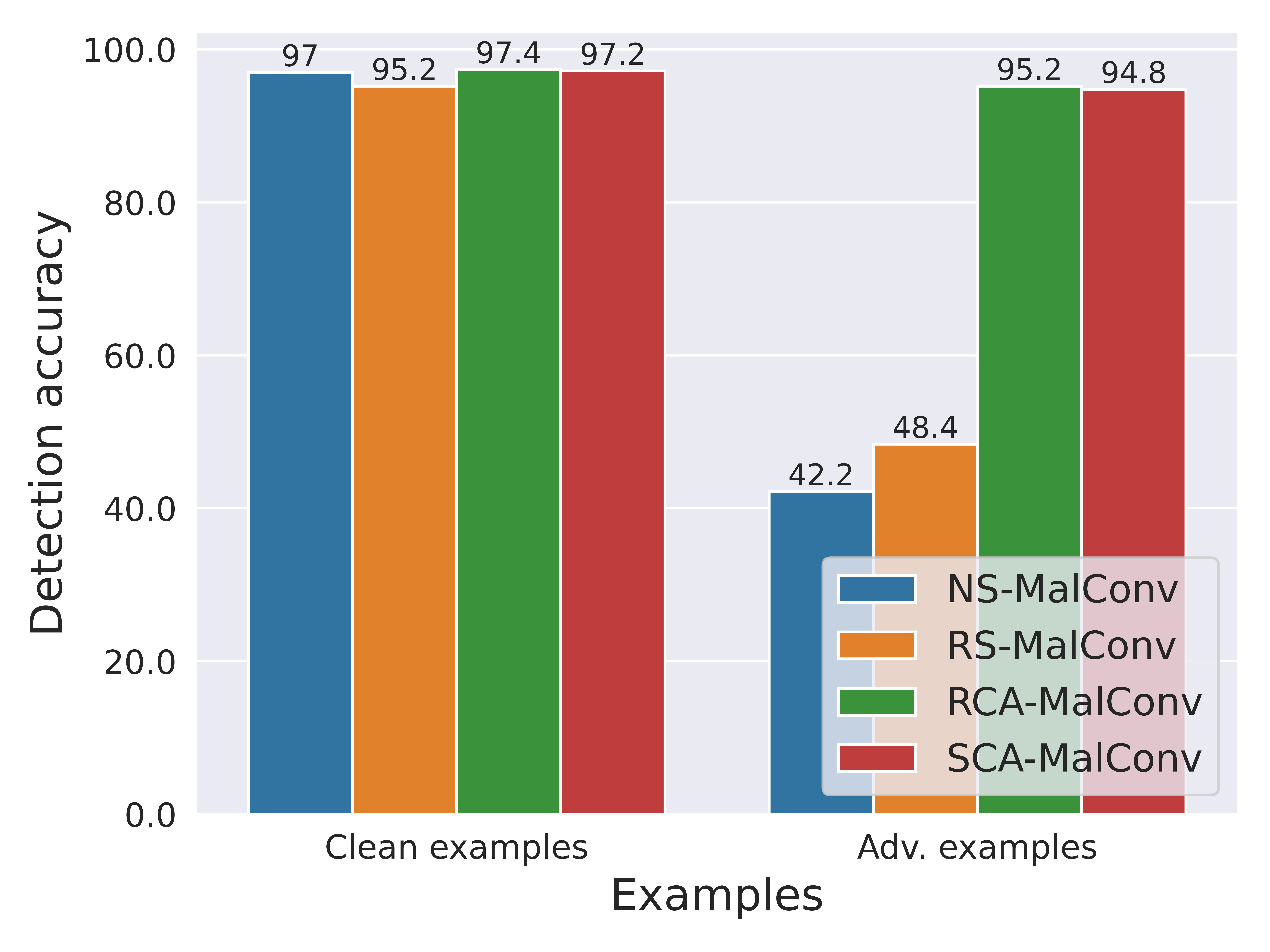}
    \caption{Detection accuracy of the malware detectors on the adversarial examples generated by the GAMMA attack~\cite{demetrio2021functionality}.}
    \label{fig:gamma_attack}
\end{figure}

Demetrio et al.~\cite{demetrio2021functionality} proposed an evasion attack that rely on injecting benign content either at the end of the file, or within some newly created sections. Afterwards, the injected adversarial payload is optimized using genetic algorithms. Figure~\ref{fig:gamma_attack} presents the detection accuracy of the ML-based models against the GAMMA attack using different hyperparameters. GAMMA allows you to define the population size, the total number of iterations to perform, i.e. 100 by default, whether to use a hard or soft label, and the amount of goodware sections to inject. In our experiments, a constraint has been imposed on the maximum size of the adversarial malware examples generated by the GAMMA attack, limiting it to twice the size of the original input. The rationale behind this decision is to mitigate the possibility of injecting excessively large adversarial payloads in pursuit of evading the malware classifier.

In Figure~\ref{fig:gamma_attack}, you can observe the adversarial accuracy of the malware detectors against the adversarial examples generated by GAMMA by injecting 10 sections and using soft label for the optimization process.
Unlike the previous of attacks, GAMMA is able to generate some adversarial examples that evade our smoothed detectors. This is because it ends up injecting huge amounts of benign content, flipping the prediction of the smoothed detectors. Nevertheless,the smoothed models are more resilient to GAMMA in comparison to the non-smoothed model. For instance, the adversarial accuracy of SCA-MalConv and RCA-MalConv is 94.80\% and 95.20\% while NS-MalConv only detects 42.20\% of the adversarial examples generated by injecting 10 sections.


\subsubsection*{Optimization of Code Caves Attack}
\begin{figure}[ht]
    \centering
    \includegraphics[width=0.6\columnwidth]{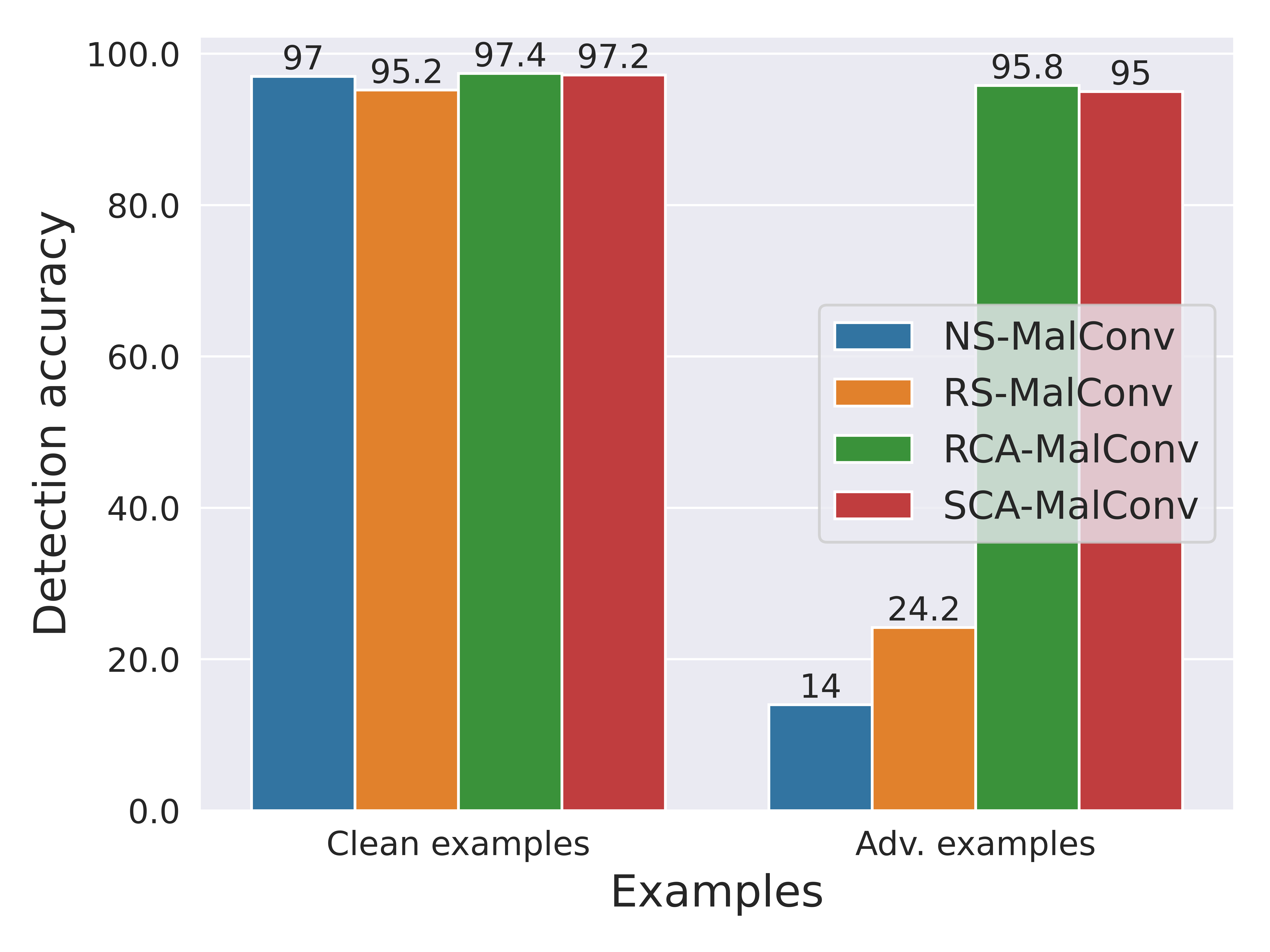}
     \caption{Detection accuracy of the malware detectors on the adversarial examples generated with the code caves optimization attack~\cite{YUSTE2022102643}. Hyperparameters of the genetic algorithm: (1) Probability of a solution in the population to be mutated, $p_1=0.1$, and (2) percentage of genes in each solution to be mutated, $p_2 = 0.1$.}
    \label{fig:code_caves_optimization_attack}
\end{figure}

Yuste et al.~\cite{YUSTE2022102643} introduced a method for generating malware examples based on dynamically extending unused blocks within the executables, referred to as code caves. Code caves are areas within the binary that are typically unused or contain extraneous code, typically employed by malware authors to embed malicious payloads with the goal of achieving misclassification. After extending the code caves, an adversarial payload is injected and optimized using genetic algorithms. Figure~\ref{fig:code_caves_optimization_attack} showcases the detection accuracy of the detectors against the code caves attack. You can observe that our chunk-based smoothing classifiers achieves notably higher detection accuracy compared to its non-smoothed counterpart, setting a new benchmark in terms of detection accuracy and representing a substantial advancement in the field. 

\subsection{Discussion}
In this section we discuss the advantages and limitations of non-smoothed and chunk-based smoothed detectors at training and test time.

The main advantage of training a malware detector on the whole binary information is that it provides a comprehensive and complete representation of the input sample. By taking as input the whole binary file, the classifier can learn to recognize patterns related to the overall structure of the binary file, including the headers, sections, segments, and specific data blocks. However, existing literature~\cite{DBLP:journals/corr/abs-1901-03583} has shown that the MalConv model does not extract meaningful characteristics from the content of the data and text sections within binary files. Instead, it tends to extract features based on attributes present in the file headers. 
On the other hand, the proposed chunk-based classifiers do not have access to a complete picture
of the binary file but only to a subset of the
bytes, without any information about the location of the chunk
of bytes in the binary file. 
By only using a
subset of the bytes to train a malware detector the computational
cost of the training process is greatly reduced, as the model only
needs to process a small portion of the input file, i.e. 5\%. Additionally, training on a portion of the file inherently introduces regularization during the training process.

At inference time, the non-smoothed detectors make predictions faster and with less computational resources than the smoothed detectors as they only assess a given input sample once. On the other hand, smoothed detectors (RCA-MalConv and SCA-MalConv) are slightly slower because they need to assess each chunk independently and then aggregate the predictions. Cf. Table~\ref{tab:models_computational_time}. However, the non-smoothed detectors are vulnerable to attacks that manipulate the entire input sample, such as injecting small adversarial payloads~\cite{DBLP:conf/sp/SuciuCJ19,demetrio2021adversarial}. On the contrary, smoothed detectors can be more accurate on adversarial examples as the adversarial code won't affect all the chunks, but a subset of them. According to the findings presented in Section~\ref{sec:nonadversarial_evaluation} and~\ref{sec:sota_attacks}, SCA-MalConv and RCA-MalConv have similar detection accuracy on clean examples as NS-MalConv while providing higher detection rates for adversarial examples. Considering these findings, it is worthwhile to consider using SCA-MalConv and RCA-MalConv over NS-MalConv for the malware detection task as the benefits of higher detection rates for adversarial examples outweigh the slightly higher inference time.

\begin{table}[ht]
\centering
\caption{Training and testing time comparison between the smoothed and non-smoothed detectors. The training procedure for both RCA-MalConv and SCA-MalConv models is identical, and thus, their training time is the same. However, their performance differ during testing as RCA-MalConv randomly samples chunks for classification while SCA-MalConv orderly samples chunks.}
\label{tab:models_computational_time}
\begin{tabular}{c|ccc}
\hline
\multirow{2}{*}{Models} & \multicolumn{3}{c}{Computational Time}                                                                                                                                 \\ \cline{2-4} 
                        & \multicolumn{1}{c|}{\begin{tabular}[c]{@{}l@{}}Training Time \\ (minutes/epoch)\end{tabular}} & \begin{tabular}[c]{@{}l@{}}Test Time - CPU\\ (seconds/example)\end{tabular} & \begin{tabular}[c]{@{}l@{}}Test Time - GPU \\ (seconds/example)\end{tabular} \\ \hline
NS-MalConv              & \multicolumn{1}{c|}{22.06}   & 0.0971    &  0.006  \\
RS-MalConv             & \multicolumn{1}{c|}{44.95}  & 1.6526    & 0.2589   \\ 
RCA-MalConv             & \multicolumn{1}{c|}{4.10*} & 0.2302   & 0.0258 \\ 
SCA-MalConv             & \multicolumn{1}{c|}{4.10*}  & 0.2867  & 0.0232  \\\hline
\end{tabular}%
\end{table}

Furthermore, our malware classification system based on chunks is interpretable by design. By assessing the maliciousness of each chunk independently, our method facilitates a finer-grained analysis of the file. This allows to identify which specific chunks within a file exhibit malicious or benign traits. As a result, the predictions for each chunk can serve as the basis for generating a visual depiction or an overview of the file, emphasizing chunks with the higher and lower maliciousness scores.

\section{Conclusions}
\label{sec:conclusions}
In this paper, we present a robust model agnostic adversarial defense against adversarial malware examples. Building upon prior research on (de)randomized smoothing, we introduce two chunk-based ablation schemes to build robust static learning-based classifiers. 
The novel application of our chunk-based ablation scheme creates new robust models, named RCA-MalConv and SCA-MalConv, that generalize better than the non-smoothed MalConv and the randomized smoothing-based MalConv against adversarial malware examples generated with state-of-the-art evasion techniques.
Our approach establishes a new benchmark in terms of detection accuracy and represents a substantial leap forward within the field. Given the inherently adversarial nature of cybersecurity, we believe that our efforts will stimulate further research in adversarial attacks and defenses for the domain of malware detection.

\subsection{Future Work}

The methods proposed in this paper are designed to classify files by breaking them down into smaller, fixed-size chunks, and then determining whether each chunk is benign or malicious. While this approach has shown promise in accurately classifying files, it is based on the assumption that all code within a malicious executable is malicious and all code within a benign executable is benign, which may not hold true in all cases. In some cases, a malicious file might only contain a small portion of malicious code while the majority of its code is benign. For simplicity purposes, in this work, all chunks within a malicious file are labeled as malicious, but a more fine-grained labelling of the chunks might improve our approach's accuracy.


Another line of research could be investigating approaches to identify and remove the adversarial content from the executables, i.e. the sequence of bytes specifically injected to flip the prediction of the classifier.

\section*{Acknowledgements}
This project has received funding from Enterprise Ireland and the European Union’s Horizon 2020 Research and Innovation Programme under the Marie Skłodowska-Curie grant agreement No 847402 and by MCIN/AEI/10.13039/501100011033/FEDER, UE under the project PID2022-139835NB-C22. We would like to thank Cormac Doherty and UCD's Centre for Cybersecurity and Cybercrime Investigation for their support.

\section*{Data and Code Availability}
The BODMAS dataset is available to the public and the source code~\footnote{\url{https://github.com/danielgibert/derandomized_smoothing_for_malware_detection}} of our approach will be made available under the MIT License after the paper is accepted.

\bibliographystyle{unsrt}  
\bibliography{references}

\end{document}